\documentclass{ifacconf}

\usepackage{graphicx}
\usepackage{natbib}
\usepackage{amsmath}
\newtheorem{assumption}{Assumption}

\newtheorem{definition}{Definition}

\begin{document}
\begin{frontmatter}

\title{\Large \bf
	{Intersection-Traffic Control of Autonomous Vehicles using Newton-Raphson Flows and Barrier Functions}}\thanks[footnoteinfo]{This work was supported in part by the National Science Foundation under grant Nos. S\&AS-1849264, CPS-1851588, and
	by ONR Minverva under grant No. N00014-18-1-2874.}
	
	\author[First]{S. Shivam},
\author[First]{Y. Wardi},
\author[First]{M. Egerstedt},
\author [Second] {A. Kanellopoulos},
\author [Second] {K. G. Vamvoudakis}
\address[First]{School of Electrical and Computer Engineering, Georgia Institute of Technology,
Atlanta, GA 30332, USA. \\
e-mail: sshivam6@gatech.edu, ywardi@ece.gatech.edu, magnus@ece.gatech.edu.}
\address [Second] {The Daniel Guggenheim School of Aerospace Engineering, Georgia Institute of Technology, Atlanta, GA, $30332$, USA.  \\
e-mail: ariskan@gatech.edu, kyriakos@gatech.edu.}
\maketitle
 \thispagestyle{empty}
\pagestyle{empty}

%\maketitle
%\thispagestyle{plain}\pagestyle{plain}
\begin{abstract}
This paper concerns an application of a recently-developed nonlinear tracking technique to trajectory control of autonomous vehicles at traffic intersections. The technique  uses a flow version
of the Newton-Raphson method for controlling a predicted system-output to a future reference target. Its implementations are based on  numerical solutions of ordinary differential equations, and it does not specify  any particular method for computing its  future reference trajectories. 
Consequently it can use relatively simple algorithms on crude models   for computing the target trajectories,  and  more-accurate models and algorithms for trajectory control in the tight loop.  We demonstrate this point 
 at an extant predictive traffic planning-and-control method with our tracking technique. Furthermore,  we guarantee safety specifications by applying to the tracking technique  the framework of control barrier functions.

\end{abstract}
\end{frontmatter}

\section{Introduction}

In a recent   work (\cite{Wardi19}), we  proposed 
 a new approach to output tracking of dynamical systems that appears to be effective while requiring  modest computing efforts. Its underscoring technique is based on
 a standalone integrator with a variable gain, designed for stability and small tracking errors. The integrator is defined by a flow version of the Newton-Raphson method for solving algebraic equations. These equations are defined by attempting to match a predicted system's output  to a predicted value of the reference target. Furthermore,  increasing the controller's rate can stabilize the system, increase its stability margins, and reduce its tracking error even (in some cases)  if the plant-subsystem is unstable and  not of a minimum phase.

The proposed tracking technique may not be as general or powerful as some of the existing nonlinear regulation techniques like the Byrnes-Isidori regulator (\cite{Isidori90}), Khalil's high-gain observers for output regulation (\cite{Khalil98}), or Model Predictive Control (MPC) (\cite{Rawlings17}). However, the effectiveness of these methods is partially due to their computational sophistication such as, respectively, nonlinear inversions, the appropriate nonlinear normal form, and real-time optimal control. As for our tracking technique, its controller is defined by an ordinary differential equation which can be solved numerically in real time and hence, as argued in \cite{Wardi19}, may be implementable by  simple algorithms. 

Thus far, the development of the proposed technique has focused on its fundamental structure, theoretical convergence results, and various  examples including an inverted pendulum and motion control in platoons (\cite{Wardi19}). Presently our  main interest is in applications to  autonomous vehicles, and especially in trajectory control of swarms and platoons. Such problems often are addressed by MPC or related techniques;   see e.g., (\cite{Kong15, Plessen18, Kim14}) and references therein. Like MPC, our technique is based on predictive control, but it is different from MPC in several ways. It is not based on optimal control nor does it specify  a particular framework for computing future target trajectories.  Once the target trajectories  become available, it uses a fluid-flow version of the Newton-Raphson method for tracking control.  

The primary objective of this paper is to investigate how the proposed tracking technique can complement other prediction-based approaches.  To this end we consider the trajectory-control technique developed in \cite{Malikopoulos18}  for  traffic management of autonomous vehicles in urban   road-intersections. This technique is slated to optimize motion-energy consumption of each vehicle while guaranteeing safety constraints. It is in the flavor of  MPC in that it solves  optimal control problems  for computing future trajectories, but unlike MPC it does not consider rolling horizons but
a single optimal control program for each vehicle approaching an intersection.
A salient feature of this technique  is that is uses  a simple dynamic model for the vehicles, comprised of a double integrator, thereby enabling   closed-form solutions to the optimal control problems.  This gives  an efficient trajectory-computation for every vehicle, which scales well with traffic loads at the intersections. 

Our tracking technique complements the traffic control framework 
of \cite{Malikopoulos18} in the following way. 
 We first 
compute the target-trajectories of the vehicles using the simple model and formula
 derived in \cite{Malikopoulos18}, then we apply our technique to a more complicated and realistic model for tracking of the target trajectory. To this end we use a dynamic bicycle model for the vehicles' motion,  a sixth-order nonlinear  model that has been extensively used in control of autonomous vehicles (see, e.g., \cite{Kong15, Plessen18} and references therein). Furthermore, we extend the applications domain of of \cite{Malikopoulos18}  from a straight road to a curved road. Lastly, we examine the treatment of  safety constraints in the tight control loop
 by incorporating control barrier functions with the tracking technique.
 
The rest of the paper is organized as follows. Section 2  formulates the problem. Section 3 summarizes the tracking technique that will be used in the sequel.  Section 4 
presents simulation results, and Section 5 concludes the paper and outlines directions for future research.

{\it Statement of contributions.} The contribution of the present paper is twofold. The first contribution extends the framework of prediction-based nonlinear tracking in the context of trajectory control of autonomous vehicles at traffic intersections, while the second is in the incorporation of safety measures through the use of barrier functions. Regarding the first contribution, the tracking technique has been applied to the dynamic bicycle model (\cite{Shivam19a,Wardi19}), and the relevant contribution in this paper is in a proof of concept regarding the way it complements the control framework of \cite{Malikopoulos18}. Regarding safety guarantees, control barrier functions have not been applied to the tracking technique or, to our knowledge, to a dynamic bicycle model.

A reduced version of this paper will appear in the {\it Proc. 21st IFAC World Congress}, Berlin, Germany, July 12-17, 2020.

\section{Problem Formulation}

Our work is concerned with the management and control of vehicle-flows at traffic  intersections. Each one of the  roads comprising an intersection consists  of two zones: a control zone and a merging zone. The merging zone  is at the center of the intersection, where lateral accidents are possible. The control zone is a stretch of the road approaching the merging zone, where the scheduling, planning and control of vehicles' trajectories  are performed. Once a vehicle enters the merging zone its speed or lane  cannot be changed. 

Whenever a vehicle enters the control zone, a scheduler computes the time and speed at which it has to enter the merging zone based on the current and future states (positions and velocities) of all the other vehicles concurrently in the intersection. Subsequently the trajectory of the newly-arrived vehicle is computed by minimizing its projected motion energy  while maximizing the throughput at the intersection, subject to  safety and operational constraints. The safety constraints include  a minimum inter-vehicle distance and a maximum deviation from a lane-center, while  the operational constraints include bounds  on speed and acceleration.  This trajectory-planning problem is formulated as an optimal control problem which is parameterized by the states  of all the other vehicles concurrently  at the intersection, hence it is different from one vehicle to the next and consequently must be solved in real time.

The contribution of  this paper is not in the aforementioned scheduling and trajectory planning-and-control problem, but in a tracking of its computed solution. Thus, the tracking control is at a lower level than the optimal control problem, and for that we use a bicycle model for the vehicles' dynamics, which is more accurate and detailed than the double-integrator model.

The bicycle model that we use is the six-degree nonlinear system
described in \cite{Kong15}.   Its state variable is 
${x=(z_{1},z_{2},v_{\ell},v_{n},\psi,\dot{\psi})^{\top}}$,
where $z_{1}$ and $z_{2}$ are the planer position-coordinates of the center of gravity of the vehicle,  $v_{\ell}$ and  $v_{n}$ are  the longitudinal and lateral velocities,  $\psi$ is the heading of the vehicle and  $\dot{\psi}$ is its angular velocity. The input,  ${u=(a_\ell , \delta_f)^\top}$, consists of  the longitudinal acceleration  and steering angle of the front wheels, respectively, and the output, $y=(z_{1},z_{2})^{\top}$, is the  center of gravity of the vehicle. The dynamic equations are
 (see \cite{Kong15}):
\begin{align}
\dot{z}_{1}&=v_{\ell}\cos\psi-v_{n}\sin\psi,\nonumber\\
\dot{z}_{2}&=v_{\ell}\sin\psi+v_{n}\cos\psi,\nonumber\\
\dot{v}_{\ell}&=\dot{\psi}v_{n}+a_{\ell},\nonumber\\
\dot{v}_{n}&=-\dot{\psi}v_{\ell}+{2}\left(F_{c,f}\cos\delta_f+F_{c,r}\right)/{m},\nonumber\\
\ddot{\psi}&={2}\left(l_f F_{c,f}\cos\delta_f-l_r F_{c,r}\right)/{I_z}, \\ \nonumber
\end{align}
where $m$ is the mass of the vehicle, $l_f$ and $l_r$ are the  front and back axles' distances from the vehicle's center of mass, 
$I_z$ is the yaw moment of inertia,  and $F_{c,f}$ and $F_{c,r}$ are the lateral forces on the front and rear wheels.  These  forces   are  approximated by the following equations,
\begin{align*}
F_{c,f}&=C_{\alpha,f}\left(\delta_f-\tan^{-1}\big((v_{n}+l_f\dot{\psi})/v_{\ell}\big)\right),\nonumber\\
F_{c,r}&=-C_{\alpha,r}\tan^{-1}\left(({v_{n}-l_r\dot{\psi}})/{v_{\ell}}\right),\nonumber
\end{align*}
where  $C_{\alpha,f}$ and $C_{\alpha,r}$ are   the cornering stiffness of the front and rear tires, respectively.

Our tracking technique will be tested first  on a curved road, which does not quite fit in the framework of \cite{Malikopoulos18} due to its one-dimensional traffic model of motion.   Then we make a more careful examination of safety constraints which are addressed in real time by control barrier functions, and for that we use a straight road in order to highlight the effects of the safety controls. 

\section{Tracking Technique and Control Barrier Functions}
This section serves to explain   the tracking technique and the
salient features of control barrier functions that will be used in the sequel. 

\subsection{Tracking Technique based on Newton-Raphson Flow}
An extension of the material  in this subsection can be found in \cite{Wardi19}.

Consider the system depicted in Figure 1, where the reference input $r(t)$, the control signal $u(t)$, and the output $y(t)$ are in $R^m$ for a given $m=1,2,\ldots$. The task of the controller is to ensure that
\begin{equation}
    \lim_{t\rightarrow\infty}||r(t)-y(t)||<\varepsilon
    \end{equation}
    for a given (small) $\varepsilon>0$.
 
 \begin{figure}	[!t]
	\centering
	\includegraphics[width=3.0 in]{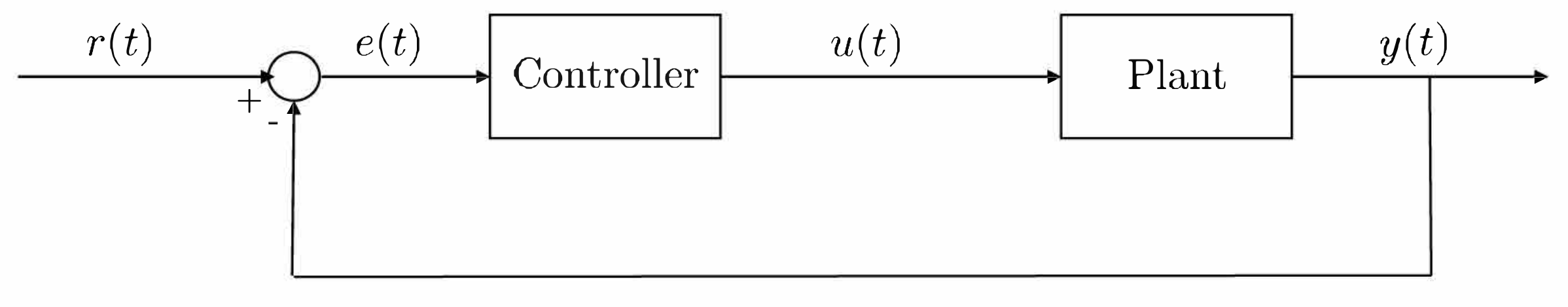}
%	\vspace{-.1in}
	\caption{{\small Basic control system.}}
\end{figure}

To explain the main idea behind the tracking technique, 
suppose first that the plant is a memoryless nonlinearity of the form
\begin{equation}
    y(t)=g(u(t))
\end{equation}
for a continuously differentiable function $g:R^m\rightarrow R^m$.  The controller that we use is defined by the differential equation
\begin{equation}
\dot{u}(t)=\alpha\Big(\frac{\textrm{d}g}{\textrm{d}u}(u(t))\Big)^{-1}\big(r(t)-g(u(t))\big),
\end{equation}
with an initial condition $u(0)\in R^m$, where $\alpha>0$ is a given speedup parameter (gain).
We implicitly assume that the Jacobian $\frac{dg}{du}(u(t))$ is nonsingular throughout the trajectory determined  by this equation. 

To see the effects of this controller on asymptotic tracking-convergence, define the Lyapunov function
\begin{equation}
    V(u(t))=\frac{1}{2}||r(t)-g(u(t))||^2.
\end{equation}
Define $\eta:=\lim\sup_{t\rightarrow\infty}||\dot{r}(t)||$. 
Then some  algebra reveals that  
\begin{equation}
\dot{V}(u(t))=\big\langle r(t)-g(u(t)),\dot{r}(t)-\alpha(r(t)-g(u(t))\big\rangle,
\end{equation}
and hence
\begin{equation}
\lim\sup_{t\rightarrow\infty}||r(t)-y(t)||~<~\frac{\eta}{\alpha};
\end{equation}
see \cite{Wardi19} for details. Stability is not a concern as long as the Jacobian $\frac{dg}{du}(u(t))$ is nonsingular throughout the trajectory $\{u(t):t\geq 0\}$ since (7) is guaranteed by (5) and (6). Therefore, $\varepsilon>0$ in (2) can be made as small as possible by taking $\alpha$ large enough in (4).

Suppose next that the plant (in Figure 1) is a dynamical system with the state equation
\begin{equation}
\dot{x}(t)=f(x(t),u(t)),\ x(0):=x_{0},
\end{equation}
and the output equation
\begin{equation}
    y(t)=h(x(t));
\end{equation}
here the state variable is $x(t)\in R^n$, $x_{0}\in R^n$ is a given initial state, and the input and output are $u(t)\in R^m$ and $y(t)\in R^m$, respectively. The dynamic response function is $f:R^n\times R^m\rightarrow R^n$, and the output function is $h:R^n\rightarrow R^m$. The following assumption, implicitly made in the forthcoming discussion, ensures that for every bounded, piecewise-continuous control function $u(t)$, and for every $x_0\in R^n$, there exists a unique continuous, piecewise continuously-differentiable solution $x(t)$ to Eq. (8).
\begin{assumption}
1). The function $f:R^n\times R^m\rightarrow R^n$ is continuously differentiable, and for every compact set $\Gamma\subset R^m$ there exists $K>0$ such that, for every $x\in R^n$ and for every $u\in\Gamma$,
\begin{equation}
\|f(x,u)\|\leq K\big( \|x\|+1\big).
\end{equation}
2). The function $h:R^n\rightarrow R^m$ is continuously differentiable.
\end{assumption}
By Eqs. (8)-(9), $x(t)$ and hence $y(t)$ are  not  functions of $u(t)$ but rather of $\{u(\tau):\tau\in[0,t)\}$ and $x_{0}$. Therefore Eq. (3) is no longer true, and a controller cannot be defined by (4). To get around this difficulty we use an output predictor and attempt to match it to a future target-reference. Given a fixed time horizon $T>0$, at every time $t\geq 0$ we compute a predictor of $y(t+T)$, denoted by $\hat{y}(t+T)$. We assume that it is a function of $x(t)$ and $u(t)$, hence has the functional
form 
\begin{equation}
    \hat{y}(t+T)=g(x(t),u(t))
\end{equation}
for a suitable function $g:R^n\times R^m\rightarrow R^m$. We henceforth implicitly assume that $g(x,u)$ is continuously differentiable.
The future reference $r(t+T)$ may have to be estimated as well by a suitable predictor (e.g., \cite{Wardi19}),  but we assume here that it is known exactly at time $t$ in order to simplify the discussion. 

The objective of the controller, next defined, is to have $\hat{y}(t+T)$ track $r(t+T)$. Thus, in analogy with (4), it is defined by the following
equation,
\begin{equation}
    \dot{u}(t)=\alpha\Big(\frac{\partial g}{\partial u}(x(t),u(t))\Big)^{-1}\big(r(t+T)-g(x(t),u(t))\big)
\end{equation}
with  given $\alpha>0$ and  initial condition $u_{0}:=u(0)$;  we implicitly assume that the partial Jacobian 
$\frac{\partial g}{\partial u}(x(t),u(t))$ is nonsingular for every $t\geq 0$. By (11), it can be seen that this controller attempts to have $\hat{y}(t+T)$ match $r(t+T)$. 

The output predictor that we use is based on the state equation (8) in the interval $\tau\in[t,t+T]$ with the input control $u(\tau)\equiv u(t)$ and initial state $x(t)$. Formally, denoting by $\xi(\tau)$ the  variable 
representing the predicted state, it satisfies the equation
\begin{equation}
    \dot{\xi}(\tau)=f(\xi(\tau),u(t)),~~~~~~~\xi(t)=x(t),
\end{equation}
then we define 
\begin{equation}
    \hat{y}(t+T)=h(\xi(t+T)).
\end{equation}
We typically solve Eq. (13) by the Forward Euler method. The following discussion implicitly assumes that this is the predictor used.

The state equation (8) and control equation (12) together define a dynamical system with input $r(t)$ and state variable $z(t):=(x(t)^{\top},u(t)^{\top})^{\top}\in R^{n+m}$. Suppose that the initial state $z_{0}:=z(0)$ is confined to a given compact set $\Gamma\subset R^{n+m}$. Ref. \cite{Wardi19} defines the following notion of BIBS stability, 
called $\alpha$-stability.
\begin{definition}
The system is $\alpha$-stable if there exist $\bar{\alpha}\geq 0$ and two class-${\mathcal K}$ functions, $\beta(s)$ and  $\gamma(s)$ such that, for every initial state $z_{0}\in\Gamma$ and an input   $\{r(t)\}$,   for every  $\alpha\geq\bar{\alpha}$, 
\begin{equation}
    ||z(t)||\leq\beta(\|z(0)\|)+\gamma(||r||_{\infty}),
\end{equation}
where $||r||_{\infty}$ is the $L^{\infty}$ norm of $\{r(t):t\geq 0\}$.
\end{definition}
Essentially $\alpha$-stability implies BIBS stability for large gain $\alpha$. Ref. \cite{Wardi19} derives verifiable sufficient conditions for $\alpha$ stability of linear systems, which cover situations where the plant subsystem is neither stable nor of a minimum phase. Simulation and experimental results of various linear and nonlinear systems indicate stability at large gains. Also, a theoretical result guarantees an extension of Eq. (7) from  the case where the plant is a memoryless nonlinearity to the present case of a dynamical system. It states that if the system is $\alpha$ stable then there exist $\eta>0$ and $\bar{\alpha}\geq 0$ such that, for every $\alpha\geq\bar{\alpha}$,
\begin{equation}
    \lim\sup_{t\rightarrow\infty}||r(t+T)-\hat{y}(t+T)||<\frac{\eta}{\alpha}.
\end{equation}
In this case, speeding up the controller by increasing $\alpha$ can serve the dual purpose of stabilizing the closed-loop system and reducing the tracking error. We point out, though, that ``tracking'' means here that the future  target reference, $r(t+T)$, is approached by the predicted output, $\hat{y}(t+T)$, not the actual output,  $y(t+T)$. In fact, defining the asymptotic prediction error by 
\[
\mu:=\lim\sup_{t\rightarrow\infty}||y(t)-\hat{y}(t)||,
\]
Eq. (7) is extended to
\begin{equation}
    \lim\sup_{t\rightarrow\infty}||r(t+T)-y(t+T)||<\mu+\frac{\eta}{\alpha}.
\end{equation}
We see that speeding up the controller does not reduce the effects of the asymptotic prediction error on the asymptotic  tracking  error. This comes at no surprise since the prediction error is akin to a measurement error in classical control systems.

Finally, a word must be said about the relationship between this tracking-control method and MPC. MPC arguably is the most-commonly used control technique based on  prediction. It solves optimal control problems in real time, and these serve the dual purpose of computing the future trajectory and controlling the system. In contrast, our technique does not specify how to compute the future target trajectory. In principle it may be given a-priori, computed once at the start of the control action as in \cite{Malikopoulos18} and in this paper, or computed on-line by data interpolation (as in \cite{Wardi19}) or neural nets (see \cite{Shivam19a}). The control law we use is based on a real-time numerical solution of a differential equation but not on optimal control. 

\subsection{Control Barrier Functions}
The authors of \cite{Ames14}  laid  the groundwork for ensuring safety   in the design of control systems. The approach combines control barrier functions with control Lyapunov functions to achieve the dual purpose of effective control and satisfying hard safety constraints.   Shortly thereafter it was applied to the control of multi-agent systems and networks, with applications to mobile robots and autonomous vehicles; see  \cite{Wang17}
for an initial work, and \cite{Ames19} for a recent survey. The essential elements of this approach are summarized in this subsection, and more-extensive expos\'es  can be found in \cite{Ames14,Ames19,Wang17}.

Consider a dynamical system defined by Eq. (8), where $x\in R^n$ and $u\in R^m$, and suppose that Assumption 1 is in force.
Let ${\cal S}\subset R^n$ be a closed set called a
{\it safe set}, and suppose that it is desirable to design a feedback control such that ${\cal S}$ is forward invariant and asymptotically stable for the system. This requirement means that (i) if $x(t_{0})\in{\cal S}$ for some $t_{0}\geq 0$, then $x(t)\in{\cal S}$ for every $t\geq t_{0}$; and (ii) the following limit holds,
\begin{equation}
    \lim_{t\rightarrow\infty}{\rm dist}\big(x(t),{\cal S}\big)=0,
    \end{equation}
    where ${\rm dist}(x,{\cal S})$ is the point-to-set Euclidean distance.
    
A continuously-differentiable function $h:R^n\rightarrow R$ is said to be a {\it barrier function} for ${\cal S}$ if there exists a Lipschitz-continuous, extended class $K$ function   $\kappa:R\rightarrow R$ (monotone increasing, $\kappa(0)=0$) such that $h(x)<0$ for every $x$ outside ${\cal S}$, $h(x)>0$ for every $x$ in the interior of ${\cal S}$, and $h(x)=0$ for all $x$ in the boundary of ${\cal S}$; and for every $t\geq 0$,
\begin{equation}
   \frac{\textrm{d}}{\textrm{d}t}h(x(t))+\kappa\big(h(x(t))\big)\geq 0. 
\end{equation}
Now suppose that a control $u(t)$ is determined such that, at each time $t\geq 0$, Eq. (19) is satisfied. Then the set ${\cal S}$ is forward invariant and asymptotically stable for the system. 
 That is, the state trajectory is driven towards ${\cal S}$, and once entering,   cannot escape from it.
The main question now is how to design a controller guaranteeing (19) for all $t\geq 0$. 

Let $u^\star(t)$ be a state-feedback control law that has been designed for a suitable performance (e.g., tracking) but without regard for safety. For every $x\in R^n$, denote by $S(x)$ the set 
\begin{equation}
S(x)=\big\{u\in R^m~:~\frac{\partial h}{\partial x}(x)f(x,u)+\kappa(x)\geq 0\big\}.
\end{equation}
Note that $\frac{\textrm{d}}{\textrm{d}t}h(x(t))=\frac{\partial h}{\partial x}(x)f(x,u)$. Therefore, if $u^\star(t)\in S(x(t))$ for every $t\geq 0$, then the requirements for the safety set ${\cal S}$ are met. However, in the event that $u^\star(t)\notin S(x(t))$ for some $t\geq 0$, it is reasonable to modify the control from $u^{\star}(t)$ to $u(t)$, defined as follows:
\begin{equation} \label{eq:quadratic_cost}
u(t)~:=\rm argmin\{||u-u^\star(t)||^2~:~u\in S(x(t))\}.
\end{equation}
In other word,
$u(t)$ is the point in $S(x)$ closest to $u^\star(t)$. With the feedback control defined by (21) for all $t\geq 0$, the requirements of the safety set ${\cal S}$ are met, and it is hoped that the performance according to which $u^\star(t)$ was defined is changed in a minimal way.

Observe that this procedure requires an optimization problem to be solved at each $t\geq 0$. 
However, if the system's dynamic equation is control-affine, then this optimization problem is a quadratic program for which there are efficient computational techniques. To see this point, suppose that 
\begin{equation}
    f(x,u)=f_{1}(x)+f_{2}(x)u
\end{equation}
for functions $f_{1}:R^n\rightarrow R^n$ and $f_{2}:R^n\rightarrow R^{n\times m}$, which means that the plant system is control affine. Then, by (21), given $t\geq 0$,    $x(t)\in R^n$ and $u^{\star}(t)\in R^m$, 
$u(t)$ is the solution of the optimization problem
\begin{eqnarray}
    \min\big\{||u-u^\star(t)||^2~:\nonumber \\
    \frac{\partial h}{\partial x}(x(t))\big(f_{1}(x(t))+f_{2}(x(t))u+\kappa\big(h(x(t))\big) \geq 0\big\};
\end{eqnarray}
the constraints on $u$ are linear hence this is a quadratic-programming problem in $u$.

We mention that  the control barrier function method, in more-general setting than described above,  has had successful applications in several  areas; see, e.g., (\cite{Ames14,Ames17,Wang17,Ames19}) and references therein.\\

\section{Simulation Results}
We consider the control zone of a road approaching an intersection, and as in \cite{Malikopoulos18},  assume that vehicles do not change lanes and hence we focus on a single lane. The motion dynamics of the vehicles  follow the bicycle model discussed in Section 2 with the following parameter values as in \cite{Shivam19}:
$m=2,050~kg$, $I_z=3,344~kg\cdot m^2$, $l_f=1.105~m$, $l_r=1.738~m$, $C_{\alpha,f}=57500~N/rad$, and $C_{\alpha,r}=92500~N/rad$. 

Two experiments are conducted. In the first experiment we consider only tracking without regard to safety constraints in the tight loop. We compute the trajectories of the vehicles by the formula derived in \cite{Malikopoulos18}, then  apply the tracking technique to ensure that the computed trajectories are followed.
In the second experiment  we  define safety constraints in terms of minimum inter-vehicle distance and maximum lateral  deviations from the lane's center, and apply Control Barrier Functions (CBF)  to ensure that they are satisfied in the face of unexpected changes to traffic conditions.\footnote{Safety has been guaranteed in the trajectory-planning stage by the scheduler and the optimal control problem. Here we  consider unforeseen situations that may arise in real time.}

\subsection{Tracking Control}
We consider a road (lane)   approaching an intersection,  consisting  of a control zone of $400$m and merging zone of $30$m. It comprises a $430$~m-long,  $30^o$ segment of a circle  defined by the following equation,
\[
z_{1}^2+(z_{2}-R)^2=R^2,
\]
where $R=\frac{430}{\pi/6}=821.24$m, $z_{1}$ and $z_{2}$ are planer coordinates of points on the circle,  $z_{1}\in[0,R\sin(\pi/6)]=[0,415.62]$,
and $z_{2}\in[0,R(1-\cos(\pi/6)]=[0,110.03]$.  
 
There are 5 vehicles in the experiment. They arrive to the control zone at randomly-drawn times,    all at the same  initial speed
of 13.4~m/s and  longitudinal acceleration of  0. 
 The 
 initial heading of all the vehicles  is $0^o$ with respect to the direction of the lane, and therefore, if the control gives effective tracking,  they
 are expected to remain close to the lane's center and  maintain a  heading of near $0^o$ (with respect to the road) throughout the control zone.

The arrival times of the vehicles to the merging zone and the vehicle's trajectories in the control zone
are computed, respectively, by the scheduling procedure and  the optimal control algorithm proposed in
\cite{Malikopoulos18}.  Now it must be pointed out that that algorithm is applicable to  straight roads since it is underscored by a straight-line model of motion. Therefore, we compute the trajectories  as if the road is straight, and map the results to the curved road according to the distance travelled.    This no-longer results in minimum-energy trajectories since the dynamic vehicle-model is two-dimensional,  but the approximation errors are minor.
 
 To test the robustness of the controller with respect to modeling variations, we induce an error of 100\% in the vehicles' mass. Thus, the predictor  equation (13) uses twice the ``real'' weight of the cars which  is used in the simulations.
 
All the differential equations for the  simulation and the controller are  solved by the Forward Euler method with step-size of $\textrm{d}t=0.005$ for the simulations, and $\Delta t=0.001$ for the  controller. The controller speedup factor is set at $\alpha=100$. 

The results are shown in Figures 2-4. Figure 2  depicts the graphs of the distance (arc-length) travelled by the five vehicles through the control zone and merging zone, as functions to time. The color-coded legend indicates the order of the vehicles, where car $i$ refers to the $i$th vehicle that arrives to the control zone. 
The vehicles travel $430$m through the control and merging zones,  but we extend the graphs past their departures from the merging zone  by holding the Distance-Travelled
variables to a constant ($430$m), for ease of a better presentation. 
Apparently 
 Car 1 moves at a constant velocity. In contrast,  subsequent vehicles slow down in order to meet  the computed schedule of  entering  the merging zone, which is more sparse than their arrival schedule to the control zone. These graphs will be used to explain some of the phenomena indicated in the figures below.

\begin{figure}	[!t]
	\centering
	\includegraphics[width=3.0 in]{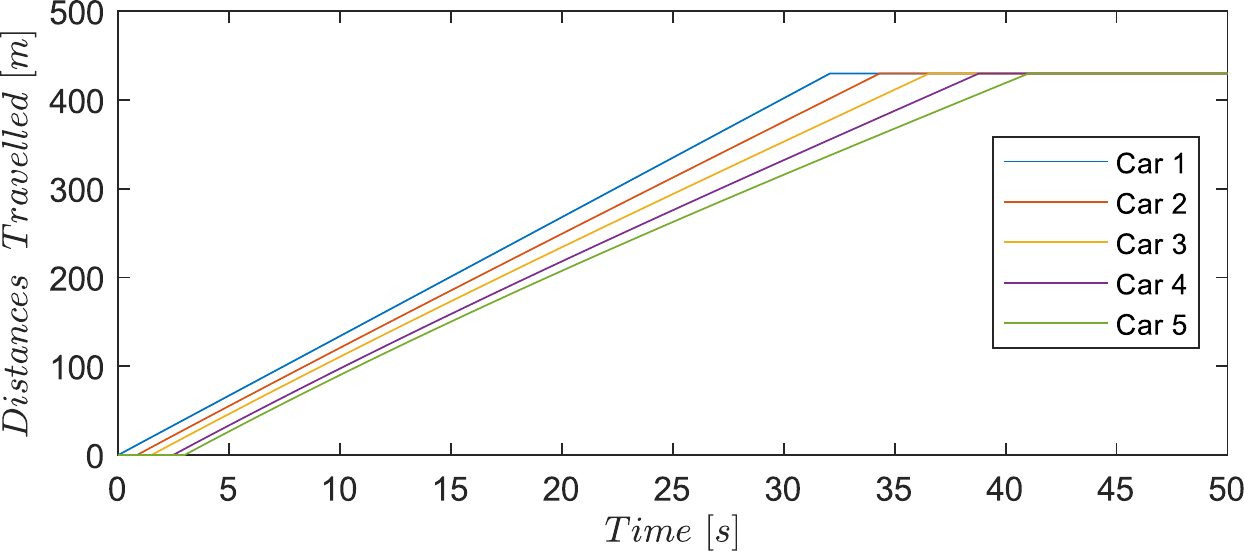}
%	\vspace{-.1in}
	\caption{{\small Distances traveled vs. time, barely distinguishable from the corresponding target trajectories. }}
\end{figure}
\begin{figure}	[!t]
	\centering
	\includegraphics[width=3.0 in]{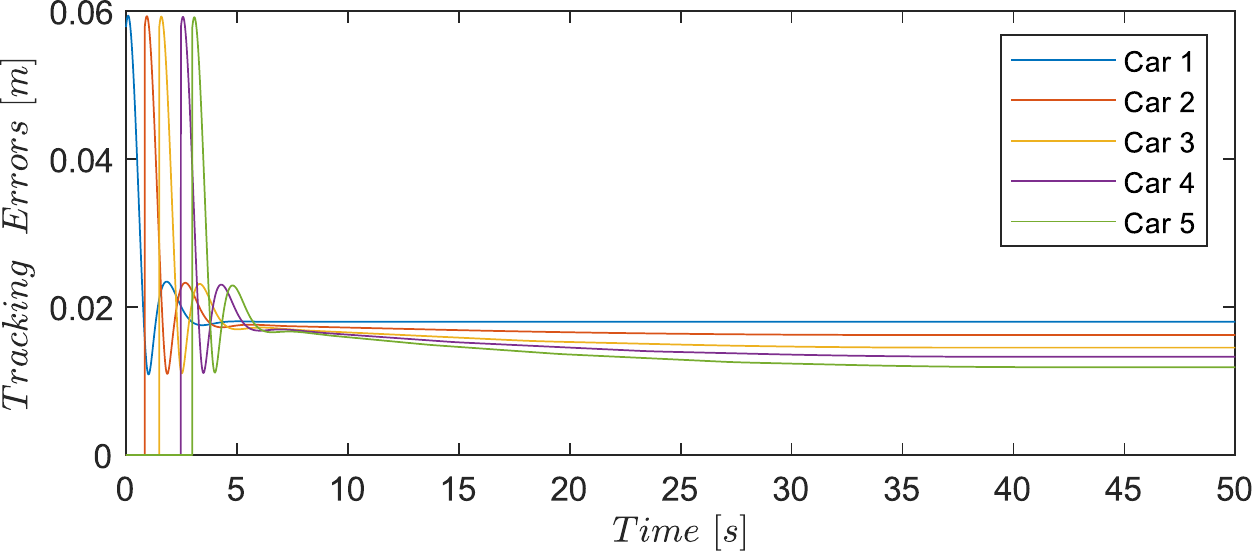}
%	\vspace{-.1in}
	\caption{{\small Tracking errors vs. time, under 6~cm during initial transient phase, and under 2~cm thereafter. }}
\end{figure}

\begin{figure}	[!t]
	\centering
	\includegraphics[width=3.0 in]{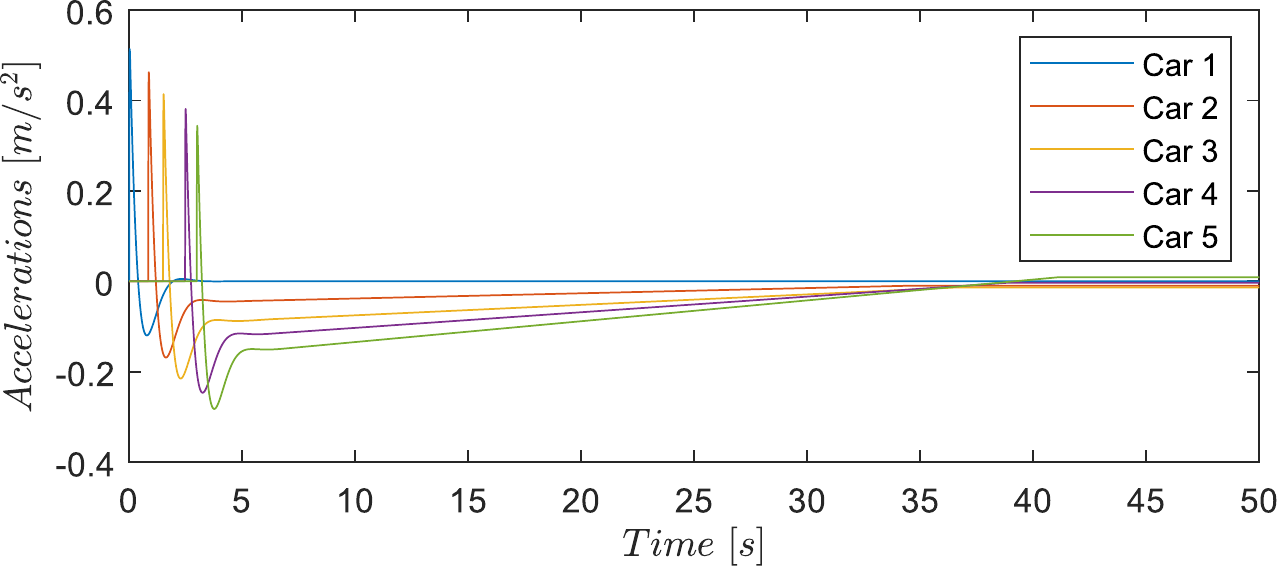}
%	\vspace{-.1in}
	\caption{{\small Vehicles' longitudinal accelerations vs. time, under 5\% of $g$. }}
\end{figure}

The tracking error for each vehicle, defined by the Euclidean distances between its position and target reference  as   computed by the optimal control program, is depicted in Figure 3. Following an initial error of about $6$~cm, the vehicles settle, in about $3$ seconds, to a steady-state error of under $2$~cm. The initial error is due to transients associated with  discrepancies between the vehicles' initial poise and their corresponding reference points.  These transients appear  identical for all the vehicles, because their respective initial positions, velocities and steering angles are identical at the respective times they enter the control zone.  The asymptotic  tracking errors are  due to the  prediction errors and  the curvatures of the road. We tested this hypothesis  by trying  the following modifications one at a time: (i) Increasing  the predictor's step size in (13) from $0.001$ to $0.002$, thereby increasing the prediction error. The resulting asymptotic  tracking errors are increased by approximately $50$\%, and the largest one, from under $2$~cm to about $3$~cm.  (ii)  Eliminating  the modeling error of the mass.
  This has  the effect of reducing the asymptotic tracking  errors, with the largest one from under   $2$~cm to $1.34$~cm. 
Since the tracking errors are small relative to the distance travelled, the graphs in Figure 2 are indistinguishable from   corresponding  graphs of the reference trajectories, which consequently are not presented in the paper.

Figure 4 depicts the graphs of the longitudinal accelerations of the vehicles vs. time, and we   observe  initial transients of under $0.48$~m/s$^2$. The graphs of later vehicles are below the graphs of earlier vehicles.
This is due to the fact that   all of the vehicles travel the same distance but later ones do it in more time,  hence have to decelerate more than earlier vehicles.
  We also note that each graph reaches zero acceleration at the final point of the control zone, Which is in compliance with the constraint that it must travel at a constant speed at the merging zone.

Finally, we tested the tracking algorithm on a straight road, all other parameters unchanged. The results are not shown here. Those of the distance travelled and longitudinal accelerations 
are barely distinguishable  from the graphs in Figure 2 and Figure 4, respectively. The only noticeable differences are in the   asymptotic  tracking errors, whose maximum  is reduced from around $2$~cm as indicated in Figure 3, to $1.34$~cm.

\subsection{Incorporation of Safety Constraints}
We next extend the simulation setting described in the last  subsection   to include safety constraints that may have to be addressed in real time, and apply to them control barrier functions.  To highlight the role of the CBF  we assume that the road is straight  and  consider only two vehicles in the forthcoming simulations. The
first vehicle serves only as a reference for controlling the second vehicle, hence we preset its trajectory and do not control it.  The second vehicle has the same dynamic bicycle model as in Subsection 4.1, and it is controlled by the same tracking technique described there. 

The first vehicle enters the control zone   at time $t=-5$~s, and the second vehicle  enters it  at time $t_{0}:=0$. The first vehicle travels along the straight road and its speed profile is shown in the blue graph of Figure 5. The constant velocities in the figure are 2~m/s, 1~m/s, and 2~m/s,  respectively, its deceleration commences at time $t_{1}$:=50~s, and its  acceleration starts at time $t_{2}$:=75~m. 
 We do not apply the tracking control or CBF to the first vehicle, and assume that it maintains the above speed profile while moving along the lane  without deviations. 

At the time the second vehicle enters the control zone (road), $t_{0}=0$, it is 10m behind the first vehicle.  Its  target reference trajectory is $r_{2}(t):=r(t)=(2t,0)^{\top}$, lying along the horizontal road. However, it enters the road at the initial steering  angle of 20$^{o}$, or 0.35~rad from the direction of the road.  Therefore initially the second vehicle  veers off the lane,  but is pulled back to it by the tracking control. Thereafter it stays on the lane while tracking its target trajectory. Without an application of the CBF to the second vehicle,  maximum deviation (lateral distance)  from the lane's center is about 1.6m, which practically may be unacceptable. Furthermore,
after returning to the lane,  it 
runs into 
 the first vehicle shortly after its slowdown. 

To avoid the collision and limit the lateral deviation from the lane, we  impose the following two safety constraints: (i) the second vehicle must maintain a distance of at least 5m from the first vehicle, and (ii) the lateral deviation of the second vehicle from the center-lane must not exceed 0.5m. We label these the {\it longitudinal constraint} and the {\it lateral constraint}. We design two corresponding CBF and apply them jointly with the tracking controller. We point out that the  longitudinal  dynamics of the second vehicle are  control-affine while its lateral dynamics are not control affine; see the state equation (1). Therefore the CFB for the longitudinal constraint can rely on quadratic programming for computing the control defined by  Eq. (21), while the lateral-safety control 
cannot use quadratic programming and has to be ad hoc.\footnote{We are aware of a transformation of the system that renders its state equation affine with respect to both input controls (\cite{Rajamani12}). However, we prefer to work with the current system in order to test the controller in an environment where the state equation is not control affine.  The results, presented in Figure 7, below, suggest that the CBF works well.}  We next explain  the two control barrier functions.

Let us denote the position and velocity of the $i$th vehicle, $i=1,\ 2$, by $p_{i}\in R^2$ and $v_{i}\in R^2$, respectively. Furthermore, define the relative displacement  and relative velocity of the second vehicle with respect to the first one by $\Delta p:=p_{1`}-p_{2}$ and $\Delta v:=v_{1}-v_{2}$. The purpose of the CBF is to ensure that $||\Delta p||\geq d_{0}$ for a given $d_{0}>0$ ($d_{0}=5$m in our experiments). Therefore it is tempting to define the safe set as ${\cal S}:=\{x\in R^6~:~||\Delta p||\geq d_{0}\}$, where $x$ is the state variable of the dynamic bicycle model defined in Section 2 (recall Eq.  (1) for its state equation). However, this can be problematic because if $||\Delta p||$ is nearly $d_{0}$ and $\Delta v$ projected on the direction of relative displacement is negative, it may be impossible to guarantee the forward invariance of the safe set. Furthermore, according to the definition of the state variable, the position of a vehicle is expressed in terms of its Cartesian coordinates while its velocity is characterized by its  longitudinal and lateral components, which can make it complicated to describe the safe set in simple terms.  

To get around this difficulty we use an idea, developed in \cite{Wang18}, of defining a CBF in terms of the relative velocity along  the relative displacement. Denoted by $\hat{v}$, it is defined by
\begin{equation}
\hat{v} = \big\langle\frac{\Delta p}{\|\Delta p\|}, \Delta v\big\rangle.
\end{equation}
Let $\bar{a}>0$ denote the maximum-possible longitudinal deceleration of the second vehicle, and define $k:=(2\bar{a})^{-1}$. Recall that the longitudinal acceleration is denoted by $a_{\ell}$ and it is a part of the input (see (1)). Now, a simple algebra shows that for every interval $[t,t_{1}]$ where the first vehicle has a constant velocity, if  $a_{\ell}(\tau)\equiv-\bar{a}$ then $||\Delta p(\tau)||\geq||\Delta p(t)||- k\hat{v}(t)^2$. 
Therefore, to ensure the forward invariance of the constraint set
    $\{x\in R^6:\|\Delta p\| \geq  d_0\}$,
    we impose the condition that 
    \begin{equation}
    \|\Delta p(t)\| - k\hat{v}(t)^2 \geq d_0. 
    \end{equation}
This leads us to define the barrier function  by
$h(x) = \sqrt{k(\|\Delta p\| - d_0)} - ||\hat{v}||$.
As a part of the safety control we enforce the condition
$\frac{d}{dt}h(x(t))+h(x(t))\geq 0$ $\forall t\geq 0$,
which implies that the set defined by (25) is forward invariant (see \cite{Ames14,Ames17,Ames19}).   Therefore, we consider the set defined by (25) as the safe set.  Finally, we note that the dynamic equation (1) is control  affine in the longitudinal acceleration, and hence the input control can be computed by a quadratic program.

To define the lateral CBF, we only need the lateral deviation of the vehicle from the lane's center  and its  velocity. Denote by $y$ the  lateral deviation, and let   $y_\textrm{max}$
be the maximum allowed deviation. In analogy to (25), the following condition ensures the   maximum deviation constraint,
\begin{equation}  \label{eq:h_horizontal_ineq}
    y_\textrm{max} \geq |y + k\frac{y}{|y|}\dot{y}^2|,
\end{equation}
where $k:=(2\tilde{a})^{-1}$ with $\tilde{a}$ denoting the maximum lateral acceleration. We define the safe set to be the set satisfying the inequality in (26).  Correspondingly, we define the barrier function
$h(x):=y_\textrm{max}-|y+k\frac{y}{|y|}\dot{y}^2|$, and define $\kappa(y)=\gamma h^3$ for a constant $\gamma>0$ (we chose $\gamma=15$). 
Taking derivatives and using (1), it can be seen that Eq. (19) is satisfied.

The input involved with the lateral deviation is the steering angle of the front wheels, $\delta_{f}$. Unlike the longitudinal control, the dynamic equation is not control affine in $\delta_{f}$, and hence we cannot compute the control by a quadratic program. Instead, a linear search is performed by a bisection algorithm over a set of admissible $\delta_{f}$,  which is the interval $[-\frac{\pi}{4},\frac{\pi}{4}]$,  and the closest value to the desired input (computed by the tracking controller)  which satisfies the (26) is chosen.

The simulation results are depicted in Figures 5-10. Figure 5 depicts the speeds of the first and second vehicles in blue and red, respectively. The visible initial transient of the red graph is due to the heading of the car when it enters the road. Note the delay in the slowdown of the second vehicle after the first vehicle; it is due to the fact that the second vehicle starts reducing its speed not immediately but when its distance from the first vehicle approaches the minimum of 5~m. In contrast, there is no such delay in the speedup since that would violate the minimum-delay constraints.  Figure 6 shows the graph of the inter-vehicle  distance, and we clearly see that it retains its minimum value through and following the speedup of the first vehicle. 

Figure 7 depicts the graph of the lateral (normal) deviation of the second vehicle from the lane's center, which is largely due to its initial heading of 20$^o$. We mentioned that without the lateral CBF the maximum distance is 1.6~m, and we observe that with the CBF, the maximum distance is about 0.27~m. Figure 8 shows the graph of the  distance between the position of the second vehicle and its target trajectory. Following an initial transient the vehicle catches up and tracks its target trajectory until the first vehicle slows down. It then rises during the slowdown period due to the action of the CBF. After  the first vehicle speeds up, the second vehicle cannot close down its tracking error since it is forced by the CBF to keep the inter-vehicle distance of 5~m, hence the tracking error assumes a constant value. 

The next two figures show the two controls.  Figure 9 depicts the longitudinal acceleration, and we notice jumps that are due to initial transients as well as the slowdown and speedup of the first-vehicle. Figure 10 depicts the graph of the steering angle of the second car, and it displays a transient due to the initial heading of the car. Neither figure displays surprising results.

\begin{figure}	[!t]
	\centering
	\includegraphics[width=3.0 in]{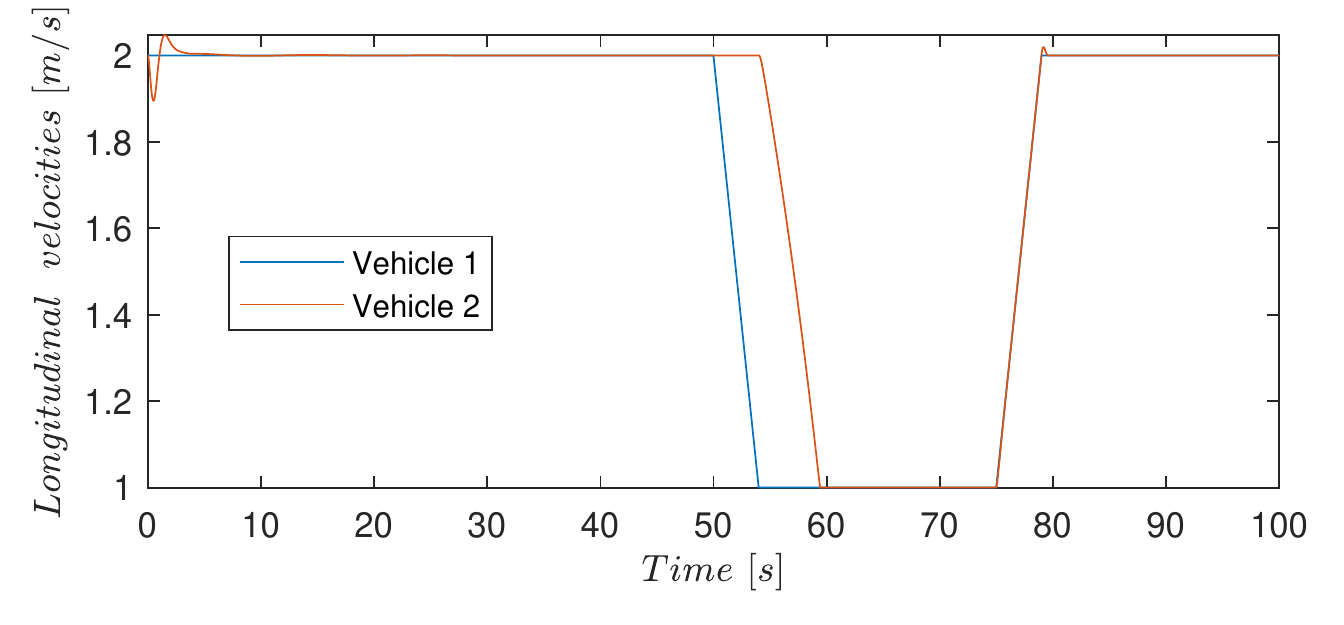}
%	\vspace{-.1in}
	\caption{{\small Vehicles' velocities vs. time. The change in velocity of Car 2 is due to the action of the CBF.}}
\end{figure}

\begin{figure}	[!t]
	\centering
	\includegraphics[width=3.0 in]{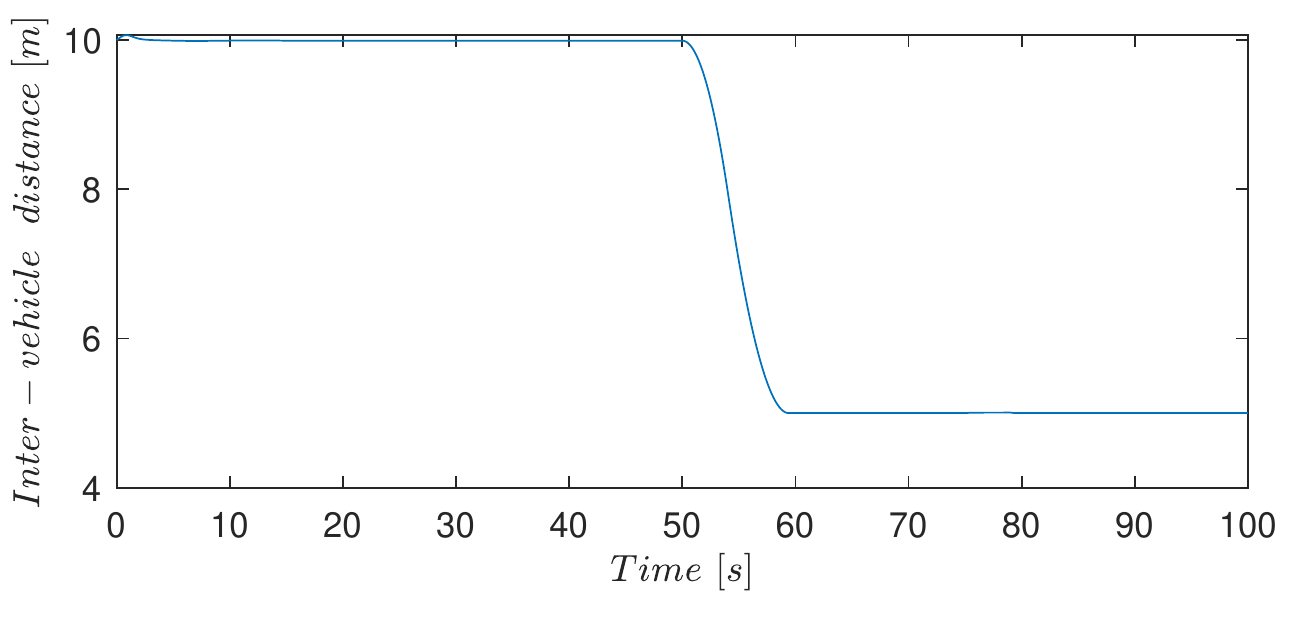}
%	\vspace{-.1in}
	\caption{{\small Distance between the two vehicles. The decline starting at 50s is due to the action of the CBF.}}
\end{figure}

\begin{figure}	[!t]
	\centering
	\includegraphics[width=3.0 in]{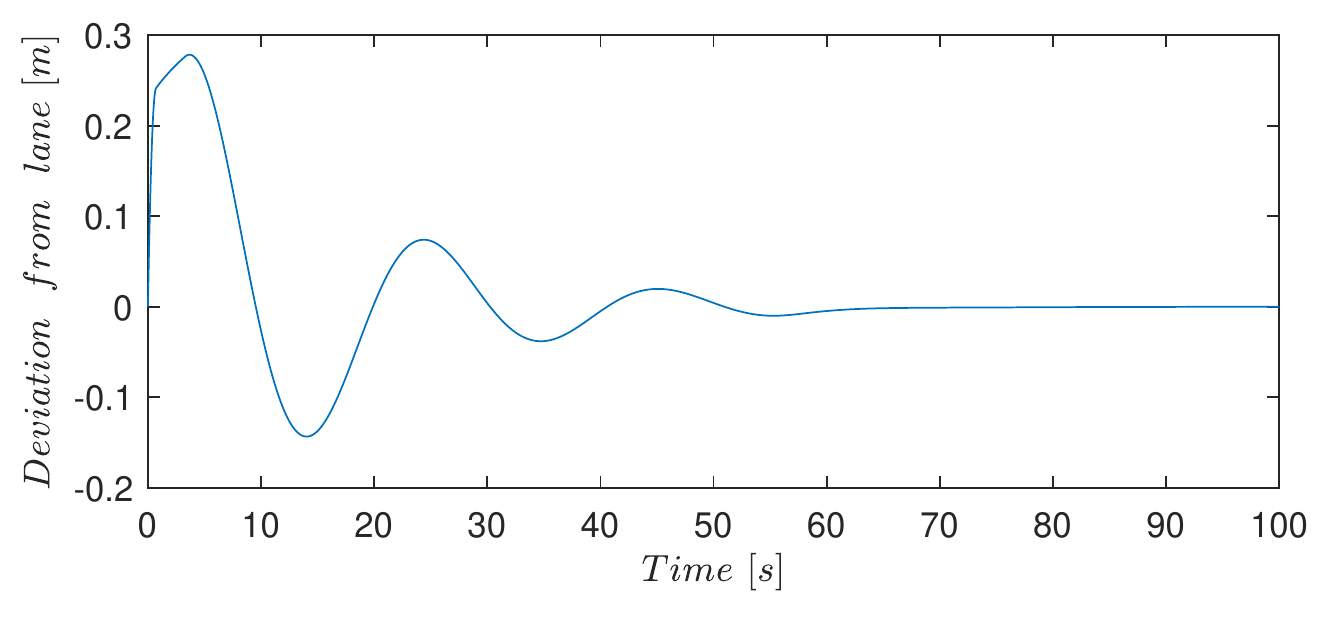}
%	\vspace{-.1in}
	\caption{{\small Distance of second vehicle from the lane-center. Without the CBF the maximum distance is about 1.6m.}}
\end{figure}

\begin{figure}	[!t]
	\centering
	\includegraphics[width=3.0 in]{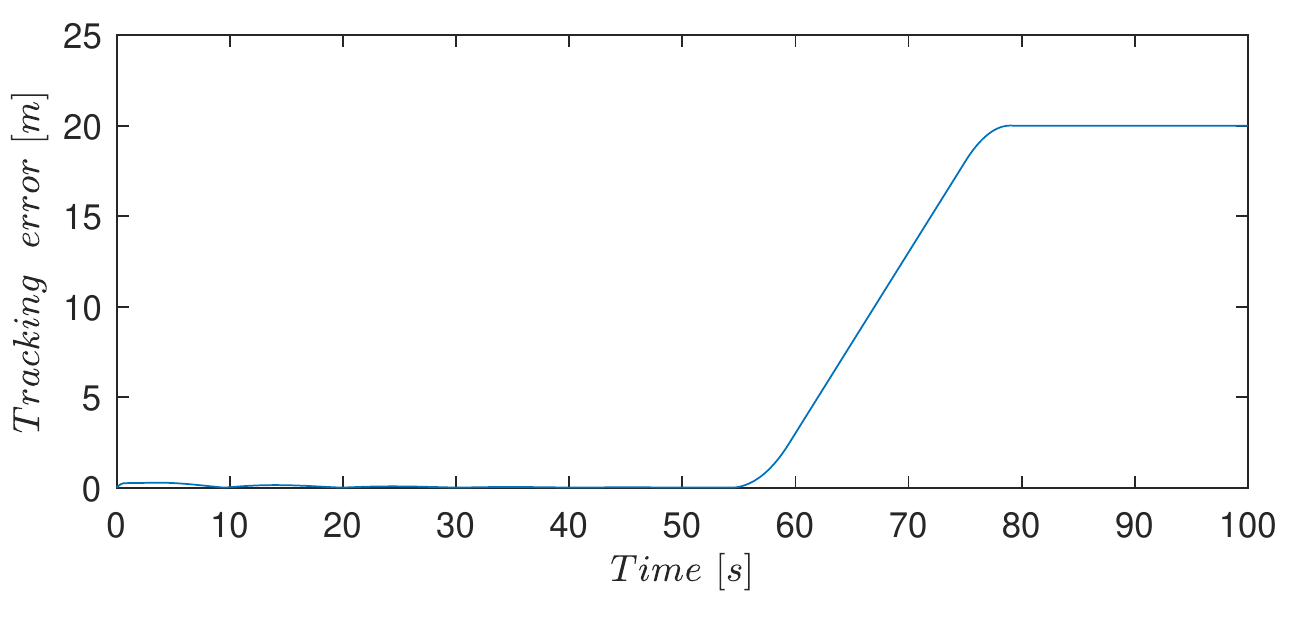}
%	\vspace{-.1in}
	\caption{{\small Tracking error of second vehicle.
	It cannot be reduced due to the CBF.}}
\end{figure}

\begin{figure}	[!t]
	\centering
	\includegraphics[width=3.0 in]{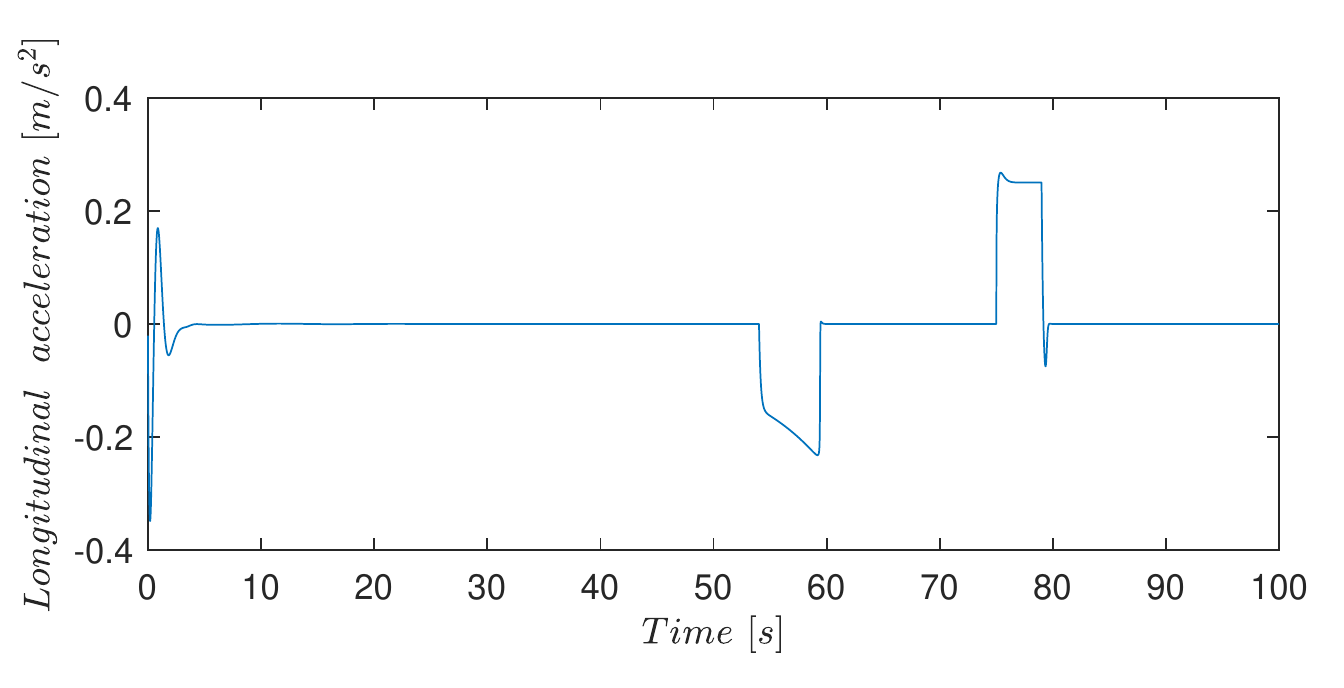}
%	\vspace{-.1in}
	\caption{{\small Longitudinal acceleration of second vehicle. }}
\end{figure}

\begin{figure}	[!t]
	\centering
	\includegraphics[width=3.0 in]{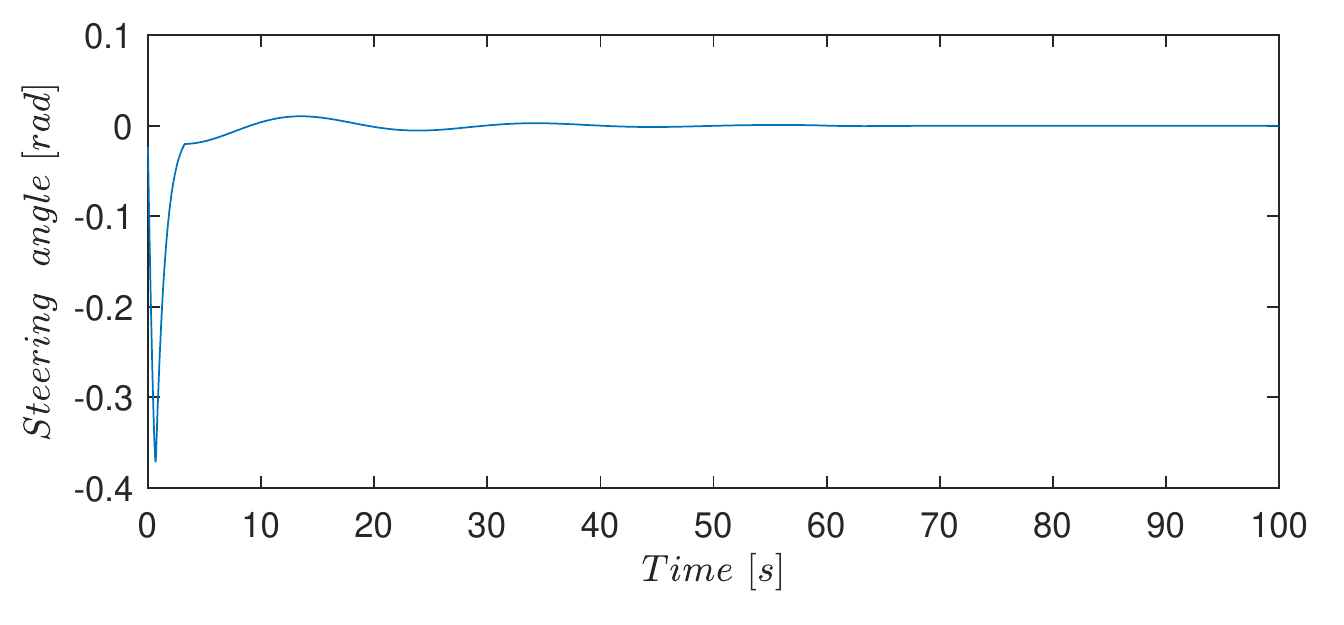}
%	\vspace{-.1in}
	\caption{{\small Steering angle of second vehicle}}
\end{figure}

\section{Conclusion and Future Work}
This work extends the framework of prediction-based nonlinear tracking in the context of trajectory control of autonomous vehicles at traffic intersections.  We present results that
use a flow version of the Newton-Raphson method for controlling a predicted system-output
to a future reference target.
Furthermore,  we  guarantee  safety  specifications  by  applying  to  the  tracking  technique  the
framework of control barrier functions.

Future work will focus on developing robustness guarantees will allow for more realistic scenarios, where noise and external disturbances are taken into consideration.

\bibliography{Paper}

\end{document}